\documentclass[aps,onecolumn,groupedaddress,floatfix,nofootinbib]{revtex4-1}

\usepackage{amsmath} 
\usepackage{amssymb}
\usepackage[version=4]{mhchem}
\usepackage{graphicx}
\usepackage{bm}
\usepackage[title]{appendix} 
\usepackage{natbib}
\usepackage[dvipsnames]{xcolor}
\usepackage[english]{babel}
\usepackage[autostyle]{csquotes}
\usepackage{algorithm} 
\usepackage{algpseudocode}
\usepackage{epigraph}
\usepackage{multirow}
\usepackage[switch]{lineno}
\usepackage[colorlinks=true,citecolor=red,filecolor=green,linkcolor=blue,pdfnewwindow=true]{hyperref} 
\usepackage{booktabs}
\usepackage{booktabs}
\usepackage{tabularx}
\usepackage{array}

\hypersetup{urlcolor=blue}

\makeatletter
 
\newcommand{\Rmnum}[1]{\expandafter\@slowromancap\romannumeral #1@}
\makeatother

\begin{document}

\title{On Optimal Control of a Genotype-Structured Prey-Predator System for Ecological Rescue and Oscillation Restoration}

\author{Preet Mishra}
\thanks{These two authors contributed equally}
\author{Shyam Kumar}
\thanks{These two authors contributed equally}
\author{Sorokhaibam Cha Captain Vyom }
\author{R.~K.~Brojen Singh}
\email{brojen@jnu.ac.in}
\affiliation{School of Computational $\&$ Integrative Sciences, Jawaharlal Nehru University, New Delhi-110067, India.}

\begin{abstract}
	
\noindent
Evolutionary changes impacts interactions among populations and can disrupt ecosystems by driving extinctions or by collapsing population oscillations, posing significant challenges to biodiversity conservation. This study addresses the ecological rescue of a predator population threatened by a mutant prey population using optimal control method. To study this, we proposed a model which incorporates genotypically structured prey comprising of wild-type, heterozygous and mutant prey types and predator population. We proved that this model has both local and global existence and uniqueness of solutions ensuring the model robustness. Then, we applied optimal control method along with Pontryagin’s Maximum Principle by incorporating a control input in the model to minimize the mutant population and subsequently to stabilize the ecosystem. The numerical results clearly reveal that the undesired dynamics of the model can be controlled showing the suppression of the mutant, rescues the predator, and restores the oscillatory dynamics of the system. These findings demonstrate the efficacy of the control strategy and provide a mathematical framework for managing such ecological disruptions. 

\end{abstract}
\maketitle

\tableofcontents


\section{Introduction}
\label{sec:introduction}

\noindent The structures of ecosystems are complex and the dynamics of these structures is shaped by interactions among the constituent species, environmental factors, and external perturbations. Shifting environmental conditions generally accelerate the creation of stressed and imbalanced diversity driven by significant change in the species interactions in the ecosystems. In ecological systems, balanced competition among species keeps diversity in equilibrium and is a very important aspect of a healthy ecosystem. However, unbalanced competition can lead to catastrophe causing drastic changes in the ecological structure.  Such disruptions may lead to a disaster outcomes, including species extinctions and the collapse of ecological oscillations, threatening biodiversity and ecosystem stability. Invasion dynamics\cite{cadotte,courchamp}, altered predator-prey systems\cite{braselton1,sun} are some of the examples, where, this imbalance in interactions have been observed in field and studied(modeled) extensively. Thus management programs working to control species population in imbalanced ecosystems \cite{courchamp} have to be equipped with adaptive strategies and proper theoretical understanding of extensive population dynamics. Also the effects of these management programs have to be properly analysed in terms of costs, budgets and performance measures\cite{kirk,lenhart}. There are extensive works on control of dynamics involving these ecological dynamics alone (for a detailed and complete bibliography, we refer to \cite{lenhart}).

\subsubsection{Brief overview of the eco-evolutionary dynamics }

\noindent In natural habitats, some ecological and evolutionary changes may be at par in time scales and may interact in a feedback loop. Examples of such eco-evolutionary dynamics in predator-prey systems are well documented. Some of the well studied field examples are included \cite{brodie1,brodie2}. These models have been extensively studied theoretically for their rich dynamics \cite{freedman,braselton1}, recent works have also been studied incorporating functional responses in predator prey systems \cite{drossel} and finally for detailed reviews we refer to the works in \cite{abrams,schoener}. The starting point of these class of models and the major underlying assumption are that without the genetically driven dynamics, we should be able to recover the usual ecological dynamical system \eqref{total_pp}.\\

\noindent  Modelling components with simplistic models by representing them with dynamical systems approach \cite{montroll}, where, the state of the ecological system at any instant of time $t$ can be represented by species population size vector $\textbf{x(t)}=[x_1,.x_i,..,x_n]^T$, where, $x_i$ is the population of the $i^{th}$ species. The dynamical behaviour of the state vector $\textbf{x}(t)$ is then given in the most general case by:

    \begin{eqnarray}
\label{dyneqn}
\frac{d\textbf{x(t)}}{dt} = \textbf{f}[{\bf x(t)},t]=\left[\begin{matrix}f_1\\f_2\\ \vdots \\f_n  \end{matrix}\right];~~\bf{x}=\left[\begin{matrix}x_1(t)\\x_2(t)\\ \vdots \\x_n(t)  \end{matrix}\right]
\end{eqnarray}
$f_i(x_1,..,x_n;t)$s can generally be inferred from experimental data and often found to be highly non-linear character \cite{May1975,Maier2013}
In this work the ecological dynamics are modeled through a two species generalized  predator-prey model with a logistic growth (with carrying capacity K) for prey and with a Monod class interaction (capture) term (also called a Hollings Type II response function) between the prey and the predator as given by :
\begin{equation}
\begin{aligned}
\frac{dx}{dt} &= \alpha x \left(1 - \frac{x}{K}\right) - \frac{mxy}{a+x} -dx\\
\frac{dy}{dt} &= y\left( \frac{mx}{a+x} - s \right).
\end{aligned}
\label{total_pp}
\end{equation}
where, $m$ is a parameter indicating the difficulty of prey capture, $a$ controls the inflection point of the functional response curve, $s$ is the intrinsic death rate of the predator,  $\alpha$ reflects the intrinsic rate of growth of prey and $K$ is carrying capacity. For the genetically driven dynamics, we use the standard one locus two allele (A and a) symmetric fertility model \cite{freedman,koth,akin}. This gives three genotypes which we keep track of i.e. AA (wild type homozygote) Aa (heterozygote) and aa (mutant). We have implicitly assumed that there is dominance of the wild type allele. These models have been well studied in \cite{freedman,koth,akin,cleve,braselton2} even showing existence of limit  cycles \cite{koth,akin}. In this work, we have taken the symmetric fertility model including differential mortality in the three genotypes \cite{koth}.\\

\noindent It is well known that equilibrium situations predicted by modeling with only ecological dynamics without evolutionary dynamics in some cases are incomplete \cite{freedman,braselton1,braselton2}. Thus, these aspects, if not taken care while designing management policies, are likely to cause serious inefficiencies in the designs. We found that optimal control problems for eco-evolutionary models have been very less studied, and hence, it is needed to extensively study such systems. Thus, in this work, we focus on the setting up of eco-evolutionary, predator-prey dynamics with optimal control problems to address some of the important issues, such as, control strategies to restore desired ecological structure, sustain such craved structure etc.\\

\subsubsection{Brief overview of optimal control}

\noindent  Optimal control theory \cite{conte,brockett1,brockett2,brockett3,neijmeier} broadly refers to the area of dynamical systems where the problem is to achieve some desired dynamic behavior by control action which is to be determined through minimizing a prescribed cost functional\cite{fleming,lenhart,kirk}. Working in this framework of optimal control in ecological models there are two very broad classes: temporal and spatio-temporal models. These models are again further sub-divided by variations within the theme. We give a small sample of the models that have been employed: dynamical systems assisted control models\cite{sun}, control of ecological model\cite{lenhart}. The fact that there is the need for multi-species models is made abundantly clear in (Page:253 in \cite{courchamp}) and in (Page:97 \cite{abrams}). These observations provides us with the rationale for choosing the minimal model which captures the above mentioned competitive dynamics of multi-species growth without making unrealistic assumptions. \\

\noindent  The primary objective of this work is to design an optimal control strategy that minimizes the mutant population while sustaining the predator and maintaining ecological oscillations, with minimal control effort possible. We formulate an optimal control problem with a cost functional that penalizes the mutant population, control effort, and regularizes the dynamics of other populations. Pontryagin’s Maximum Principle (PMP) is employed to derive necessary conditions for the optimal control, yielding a two-point boundary value problem (TPBVP) involving state and costate dynamics\cite{kirk}. Due to the system’s high nonlinearity, analytical solutions are infeasible, necessitating numerical methods to compute the control trajectory.

\subsubsection{Objectives of the work}

\noindent  The objectives of this work are:
\begin{enumerate}
    \item Developing a nonlinear extended predator-prey model capturing mutant-driven ecological disruption (eco-evolutionary dynamics) and incorporating optimal control strategy.
    \item Local and global existence proofs of the proposed model ensuring the model robustness.
    \item Deriving an optimal control strategy via Pontryagin principle (PMP) to achieve ecological rescue.
    \item Validating the control strategy through numerical simulations, demonstrating practical applicability.
\end{enumerate}

\subsubsection{ A brief overview of the paper}

\noindent  In this work, we address the ecological rescue problem of controlling a mutant population to sustain a predator species and restore oscillatory population dynamics, motivated by real-world challenges such as invasive species management and genetic threats to ecological balance. Our model system comprised of a prey which has three genotypes governed by a system of nonlinear ordinary differential equations (ODEs). To counteract this, we introduce a control input \(u(t)\), representing suppression efforts (e.g., culling or treatment) targeting the mutant population. The state vector \(\mathbf{x} = [x_1, x_2, x_3, y]^T\) evolves in a biologically constrained domain, ensuring non-negative populations and a positive total prey population.\\

\noindent  Theoretically, we establish the well-posedness of the model by proving local existence and uniqueness of solutions, ensuring the system’s mathematical validity. Additionally, we demonstrate global existence and boundedness under a parameter condition, confirming that solutions remain biologically realistic over time. Numerically, we solve the TPBVP using the CasADi framework \cite{casadi} with the CVODES integrator and IPOPT solver. Simulations compare uncontrolled and controlled dynamics: without control, the mutant population grows, causing predator extinction and oscillation collapse. However, with optimal control, the mutant population is suppressed, the predator is rescued, and oscillatory dynamics are restored.\\

\noindent  The paper is organized as follows: Section \ref{sec:methods} presents the model, optimal control problem, and numerical methods. Section \ref{sec:results} details the theoretical results (existence and uniqueness, PMP analysis, numerical simulations). Section \ref{sec:discussion} discusses implications and future directions, followed by conclusions in Section \ref{sec:conclusion}.

\newcolumntype{L}[1]{>{\raggedright\arraybackslash}p{#1}}

\begin{table*}[t]
\centering
\renewcommand{\arraystretch}{1.3} 
\begin{tabularx}{\textwidth}{@{}L{2.5cm} L{4.2cm} X@{}} 
\toprule
\textbf{Symbol} & \textbf{Variable Name} & \textbf{Description} \\
\hline
\hline
\midrule
\multicolumn{3}{@{}l}{\textbf{State and Control Variables}} \\
\hline\\
\midrule
$\mathbf{x} = [x_1, x_2, x_3, y]^T$ & State vector & Wild-type (AA)($x_1$), heterozygous(Aa) ($x_2$), mutant(aa) ($x_3$), and predator ($y$) population. $\mathbf{x} \in D \subset \mathbb{R}^4$, with $D = \{\mathbf{x} : x_i, y \geq 0, \sum_{i=1}^3 x_i \geq \epsilon\}$ \\
$u(t)$ & Control input & Suppression effort targeting mutants; $u \in [0, u_{\max}]$ \\
$\bm{\lambda}$ & Costate vector & Costates for Pontryagin’s Maximum Principle (PMP), $\bm{\lambda} \in \mathbb{R}^4$ \\
\hline
\midrule
\multicolumn{3}{@{}l}{\textbf{Model Parameters}} \\
\hline\\
\midrule
$\alpha_i$ & Growth rates & Intrinsic growth for $x_i$, $\alpha_i > 0$  \\
$\beta_{ij}$ & Fertilities & Fertilities values of various genotype matings $\beta_{ij} > 0$  \\
$K$ & Carrying capacity & Population limit, $K > 0$  \\
$m_i$ & Predation rates & Difficulty of prey capture $m_i > 0$ \\
$a$ & Half-velocity constant & Controls the half saturation value of the functional response $a > 0$  \\
$d_i$ & Mortality rates & Natural mortality depending on genotype, $d_i > 0$  \\
\hline
\midrule
\multicolumn{3}{@{}l}{\textbf{Cost Parameters}} \\
\hline\\
\midrule
$q$ & Mutant penalty & Weight on $x_3$ deviation, $q > 0$ \\
$c$ & Control cost & Weight on effort, $c > 0$  \\
$p$ & Terminal cost & Final $x_3$ penalty, $p > 0$  \\
$x_{3,\mathrm{target}}$ & Target & Desired $x_3$ level  \\
\hline
\hline\\
\bottomrule
\end{tabularx}
\caption{Summary of variables and parameters used in the ecological control model, including state variables, model parameters, cost terms, and numerical settings.}
\label{table1}
\end{table*}

\section{Model and Methods}\label{sec:methods}

\subsection{Optimal control problem for Predator-prey model involving 3 genotypes}

\noindent We model the generic predator-prey system along with the genetically driven dynamics in the prey population by a single locus, symmetric fertility model with mortality differences in genotype. The growth dynamics of the three prey genotypes wild-type(AA) (\(x_1\)), heterozygous(Aa) (\(x_2\)), mutant(aa) (\(x_3\)), and predator (\(y\)) populations, are governed by a system of nonlinear ordinary differential equations (ODEs). The dynamics are chosen such that it reflects the scenario where the mutant prey population is harmful to the prey. Thus the mutant prey, if unchecked, drives the predator to extinction by altering predation dynamics, leading to a loss of oscillatory behavior critical to ecological resilience. To counteract this, we introduce a control input \(u(t)\), representing suppression efforts (e.g., culling or treatment) targeting the mutant population. The state vector \(\mathbf{x} = [x_1, x_2, x_3, y]^T\) evolves in a biologically constrained domain, ensuring non-negative populations and a positive total prey population. The variables and parameters used in the model are provided in the table \ref{table1}.\\

\noindent A significant dynamical behavior of this model is that due to the evolution of the harmful mutant prey type(aa), there are extinction events of predator population. The situation is further aggravated by the scenario when the mutant prey becomes established at a high level it eliminates not only the predator at a very small time scale but also the wild type (AA) population are wiped out. We have factored these scenarios in our model as our goal is to study the optimal control of such scenarios to save and sustain the desired ecological structure.\\

\noindent The optimal control problem thus set up is given by the following nonlinear coupled ODEs for the growth of each of the population:
\begin{align}
\frac{dx_1}{dt}  &= \alpha_1 x \left[ \beta_{11} \left( \frac{x_1}{x} \right)^2 + \beta_{12} \frac{x_1 x_2}{x^2} + \beta_{22} \frac{1}{4} \left( \frac{x_2}{x} \right)^2 \right] - \frac{\alpha_1}{K} x_1 x  -\frac{x_1}{x} \frac{m_1xy}{a+x} -d_1x_1  \label{eq:x1} \\
\frac{dx_2}{dt}  &= 2 \alpha_2 x \left[ \beta_{13} \frac{x_1 x_3}{x^2} + \frac{1}{2} \beta_{23} \frac{x_2 x_3}{x^2} + \beta_{12} \frac{1}{2} \frac{x_1 x_2}{x^2} + \frac{1}{4} \beta_{22} \left( \frac{x_2}{x} \right)^2 \right] - \frac{\alpha_2}{K} x_2 x -\frac{x_2}{x} \frac{m_2xy}{a+x}-d_2x_2  \label{eq:x2} \\
\frac{dx_3}{dt} &= \alpha_3 x \left[ \beta_{33} \left( \frac{x_3}{x} \right)^2 + \beta_{32} \frac{x_3 x_2}{x^2} + \frac{1}{4} \beta_{22} \left( \frac{x_2}{x} \right)^2 \right] - \frac{\alpha_3}{K} x_3 x -\frac{x_3}{x} \frac{m_3xy}{a+x}-d_3x_3- u(t) x_3 \label{eq:x3} \\
\frac{dy}{dt} &= y\left[ \frac{m_1x_1+m_2x_2}{a+x} -s   \right] -\frac{m_3x_3y}{a+x}  \label{eq:y} \\
x&=x_1+x_2+x_3
\end{align}
where \(x = x_1 + x_2 + x_3\) is the total population, \(\alpha_i, \beta_{ij}, K, m_i, a, d_i, s > 0\) are strictly positive parameters, and \(u(t) \in \mathbb{R}\) is the control input. The state vector is \(\mathbf{x} = [x_1, x_2, x_3, y]^T \in \mathbb{R}^4\), and the system is written compactly as

$$\dot{\mathbf{x}}(t) = f(\mathbf{x}(t), u(t)), \quad \mathbf{x}(0) = \mathbf{x}_0$$,

\noindent with initial condition \(\mathbf{x}_0 = [x_{1_0}, x_{2_0}, x_{3_0}, y_0]^T\). The control \(u(t)\) is assumed to be measurable and bounded, i.e., \(u(t) \in L^\infty[0, T]\) with \(|u(t)| \leq u_{\max}\) for some \(u_{\max} > 0\). The domain is defined as

$$D = \{ \mathbf{x} = [x_1, x_2, x_3, y]^T \in \mathbb{R}^4 : x_i, y \geq 0, \Sigma_i x_i \geq \epsilon \}$$,

\noindent where \(\epsilon > 0 \ \forall t \) ensures a positive total population, reflecting biological constraints.

\subsubsection{Methods for Optimal Control} 
\noindent We formulate a general optimal control problem for a nonlinear dynamical system to minimize a cost functional subject to state dynamics. Consider the system:
$$\dot{\mathbf{x}}(t) = f(\mathbf{x}(t), u(t)), \quad \mathbf{x}(0) = \mathbf{x}_0$$

\noindent where \(\mathbf{x}(t) \in \mathbb{R}^n\) is the state vector, \(u(t) \in \mathcal{U} \subset \mathbb{R}^m\) is the control, and \(f: \mathbb{R}^n \times \mathbb{R}^m \to \mathbb{R}^n\) is a nonlinear function. The performance index to be minimized is:
$$J(u) = \phi(\mathbf{x}(T)) + \int_0^T L(\mathbf{x}(t), u(t)) \, dt $$
where \(\phi: \mathbb{R}^n \to \mathbb{R}\) is the terminal cost, \(L: \mathbb{R}^n \times \mathbb{R}^m \to \mathbb{R}\) is the running cost, and \(T > 0\) is the fixed time horizon.

\noindent To incorporate the dynamics as constraints, we define the augmented objective function using Lagrange multipliers \(\bm{\lambda}(t) \in \mathbb{R}^n\):
$$J_a(u, \mathbf{x}, \bm{\lambda}) = \phi(\mathbf{x}(T)) + \int_0^T \left[ L(\mathbf{x}, u) + \bm{\lambda}^T (f(\mathbf{x}, u) - \dot{\mathbf{x}}) \right] dt$$

\noindent  The Hamiltonian is:
$$\mathcal{H}(\mathbf{x}, u, \bm{\lambda}) = L(\mathbf{x}, u) + \bm{\lambda}^T f(\mathbf{x}, u)$$
Pontryagin’s Maximum Principle provides the necessary conditions for optimality:
\begin{itemize}
    \item \textbf{State Dynamics}: \(\dot{\mathbf{x}}^*(t) = \frac{\partial \mathcal{H}}{\partial \bm{\lambda}} = f(\mathbf{x}^*(t), u^*(t))\), \(\mathbf{x}^*(0) = \mathbf{x}_0\).
    \item \textbf{Costate Dynamics}: \(\dot{\bm{\lambda}}(t) = -\frac{\partial \mathcal{H}}{\partial \mathbf{x}}(\mathbf{x}^*(t), u^*(t), \bm{\lambda}(t))\), with \(\bm{\lambda}(T) = \frac{\partial \phi}{\partial \mathbf{x}}(\mathbf{x}^*(T))\).
    \item \textbf{Minimization Condition}: \(u^*(t) = \arg \min_{u \in \mathcal{U}} \mathcal{H}(\mathbf{x}^*(t), u, \bm{\lambda}^*(t))\).
\end{itemize}
These conditions define a two-point boundary value problem, solved numerically to obtain the optimal control \(u^*(t)\).

\subsubsection{Numerical Methods}
\label{sec:numerics}

\noindent To solve the two-point boundary value problem (TPBVP) from the optimal control problem, we employ the CasADi framework. The time horizon \([0, T]\) is discretized into \(N\) intervals with step size \(\Delta t = T/N\), defining state \(\mathbf{x}(t_k) \in \mathbb{R}^n\) and control \(u(t_k) \in \mathcal{U} \subset \mathbb{R}^m\) at grid points \(t_k = k \Delta t\), \(k = 0, \ldots, N\). The cost functional \(J(u) = \phi(\mathbf{x}(T)) + \int_0^T L(\mathbf{x}, u) \, dt\) is approximated via numerical quadrature. The dynamics \(\dot{\mathbf{x}} = f(\mathbf{x}, u)\) are enforced as equality constraints using the CVODES integrator from SUNDIALS. The resulting nonlinear programming problem (NLP) is formulated in CasADi and solved using the IPOPT solver with the MUMPS linear solver and tolerance \(10^{-2}\), respecting control constraints \(u \in \mathcal{U}\) and initial condition \(\mathbf{x}(0) = \mathbf{x}_0\). This yields the optimal control \(u^*(t)\).

\section{Results}\label{sec:results}

\noindent This section presents the theoretical results for the controlled population dynamics system, establishing the foundation for optimal control strategies to suppress the harmful mutant population ($x_3$) while maintaining ecological balance. We first prove the existence and uniqueness of solutions to ensure the system’s well-posedness. Then, we apply Pontryagin’s Maximum Principle (PMP) to derive necessary conditions for the optimal control, characterizing the suppression strategy.

\subsection{Existence and Uniqueness of Solutions}\label{subsec:existence_uniqueness}

\noindent In this section we prove two results :
(i)We establish the existence, uniqueness, and  (ii) global boundedness implying global existence of solutions to a controlled population dynamical system modeling of the interactions among wild-type ($x_1$), heterozygous ($x_2$), mutant ($x_3$), and predator (\(y\)) populations, with a control input \(u(t)\) to suppress the mutant population. For the proofs we have followed the method given in (Sec 6.2,pg:149-150) of \cite{strogatz} and(Sec 2.2 ,pg : 70 of \cite{perko})

\subsubsection{Local Existence and Uniqueness}

\noindent We first establish local existence and uniqueness of solutions.

\newtheorem{theorem}{Theorem}
\begin{theorem}\label{thm:existence_uniqueness}
Let \(D\) be as defined above, and let \(U = [0, u_{\max}]\). For any initial condition \(\mathbf{x}(0) = \mathbf{x}_0 \in D\) and control \(u(t) \in L^\infty[0, T]\), there exists a \(\delta > 0\) such that the system \eqref{eq:x1}--\eqref{eq:y} has a unique solution \(\mathbf{x}(t) \in C^1([0, \delta]; D)\) satisfying \(\mathbf{x}(0) = \mathbf{x}_0\).
\end{theorem}

\noindent\textbf{Proof:}
\noindent We apply the Picard-Lindel\"{o}f theorem for nonlinear ODEs with control inputs. \\
Define \(f(\mathbf{x}, u) = [f_1(\mathbf{x}), f_2(\mathbf{x}), f_3(\mathbf{x}, u), f_4(\mathbf{x})]^T\), where \(f_1, f_2, f_3, f_4\) are the right-hand sides of \eqref{eq:x1}--\eqref{eq:y}. We verify:
\begin{enumerate}
    \item \textbf{Continuity}: \(f(\mathbf{x}, u)\) is continuous in \(\mathbf{x} \in D\) and \(u \in U\).
    \item \textbf{Lipschitz continuity in \(\mathbf{x}\)}: For each \(u \in U\), there exists \(L > 0\) such that \(\| f(\mathbf{x}_1, u) - f(\mathbf{x}_2, u) \| \leq L \| \mathbf{x}_1 - \mathbf{x}_2 \|\) for all \(\mathbf{x}_1, \mathbf{x}_2 \in D\).
    \item \textbf{Admissible control}: \(u(t) \in L^\infty[0, T]\) is measurable and bounded.
\end{enumerate}

\noindent 1. \textit{Continuity:}
Consider a bounded subset \(D_b = \{ \mathbf{x} \in D : 0 \leq x_1, x_2, x_3, y \leq M, x_1 + x_2 + x_3 \geq \epsilon \}\), where \(M > 0\) is chosen such that \(\mathbf{x}_0 \in D_b\). For \(f_1\):

\[
f_1(\mathbf{x}) = \alpha_1 x \left( \beta_{11} \frac{x_1^2}{x^2} + \beta_{12} \frac{x_1 x_2}{x^2} + \beta_{22} \frac{x_2^2}{4 x^2} \right) - \frac{\alpha_1}{K} x_1 x - \frac{x_1 m_1 x y}{a + x} - d_1 x_1.
\]

\noindent The growth term involves fractions \(\frac{x_i x_j}{x^2}\), which are continuous since \(x \geq \epsilon\). The carrying capacity term \(-\frac{\alpha_1}{K} x_1 x\) is polynomial, the predation term \(-\frac{x_1 m_1 x y}{a + x}\) is continuous as \(a + x \geq a > 0\), and the death term \(-d_1 x_1\) is linear. At \(y = 0\), the predation term is zero, preserving continuity. Similarly, \(f_2\) and \(f_3\) involve rational functions with denominators \(x^2 \geq \epsilon^2\) and \(a + x \geq a\), and the control term \(-u x_3\) in \(f_3\) is continuous since \(|u| \leq u_{\max}\). For \(f_4\):

\[
f_4(\mathbf{x}) = y \left( \frac{m_1 x_1 + m_2 x_2}{a + x} - s \right) - \frac{m_3 x_3 y}{a + x},
\]

\noindent the fractions are continuous, and multiplication by \(y\) ensures continuity at \(y = 0\). Thus, \(f(\mathbf{x}, u)\) is continuous on \(D_b \times U\).

\noindent 2. \textit{Lipschitz Continuity}
We show that all the terms in the Jacobian \(\frac{\partial f}{\partial \mathbf{x}}\) are bounded on \(D_b\).

\noindent For \(f_1\), let \(g_1 = \beta_{11} \frac{x_1^2}{x^2} + \beta_{12} \frac{x_1 x_2}{x^2} + \beta_{22} \frac{x_2^2}{4 x^2}\). The partial derivative is:
$$\frac{\partial f_1}{\partial x_1} = \alpha_1 g_1 + \alpha_1 x \left( \beta_{11} \frac{2 x_1 x^2 - x_1^2 \cdot 2 x}{x^4} + \beta_{12} \frac{x_2 x^2 - x_1 x_2 \cdot 2 x}{x^4} \right) - \frac{\alpha_1}{K} (x + x_1) - m_1 y \frac{(a + x) x - x_1 x}{(a + x)^2} - d_1 $$

\noindent Since \(x \geq \epsilon\), \(x_i, y \leq M\), and \(a + x \geq a\), each term is bounded. For example, \(\frac{2 x_1 (x - x_1)}{x^3} \leq \frac{2 M \cdot 3 M}{\epsilon^3}\). Similarly, \(\frac{\partial f_1}{\partial x_2}, \frac{\partial f_1}{\partial x_3}, \frac{\partial f_1}{\partial y}\) involve rational functions bounded on \(D_b\).Since all the terms are in same form we have shown the continuity in case of one term the rest terms follow analogously. For \(f_4\):

\[
\frac{\partial f_4}{\partial y} = \frac{m_1 x_1 + m_2 x_2}{a + x} - s - \frac{m_3 x_3}{a + x} \leq \frac{m_1 M + m_2 M}{a} + s + \frac{m_3 M}{a},
\]

\noindent which is finite. The control term in \(f_3\), \(-u x_3\), has \(\frac{\partial}{\partial x_3} (-u x_3) = -u \leq u_{\max}\). Thus, \(\frac{\partial f}{\partial \mathbf{x}}\) is bounded, implying Lipschitz continuity with constant \(L > 0\).

\noindent 3. \textit{Admissible Control}
The control \(u(t) \in L^\infty[0, T]\) with \(|u(t)| \leq u_{\max}\) is measurable and bounded.

\noindent By the Picard-Lindel\"{o}f theorem, there exists \(\delta > 0\) such that a unique solution \(\mathbf{x}(t) \in C^1([0, \delta]; D_b) \subset C^1([0, \delta]; D)\) exists. Hence the theorem is proved.

\subsubsection{Global Existence and Boundedness}

\noindent We now extend the solution to global existence, assuming the control is non-negative, consistent with its role in suppressing the mutant population.

\begin{theorem}\label{thm:global}
Under the assumptions of Theorem \ref{thm:existence_uniqueness}, with \(u(t) \in [0, u_{\max}]\) and the parameter condition \(\max(m_1, m_2) X_{\max} < s a\), where \(X_{\max}\) is given by  $\max(X(0), \frac{K [C \max_i (\alpha_i) - \min_i (d_i)]}{\min_i (\alpha_i)})$ , the solution \(\mathbf{x}(t)\) exists on \([0, \infty)\) and remains bounded in \(D\).
\end{theorem}

\noindent  \textbf{Proof: }\\
\noindent We proceed in three steps: non-negativity, boundedness of the total prey population, and boundedness of the predator population thus guaranteeing the global existence. There are three types of terms involved in the vector fields : (i) growth terms (ii) carrying capacity terms and (iii) predation,death and control terms . We check the growth estimates for each of the term and prove that they do not blow up in finite time, thus giving us boundedness. Further since we have already proved that the functions involved in the vector fields are Lipschitz and rational functions this implies global existence.

\noindent\paragraph{Non-negativity}
The non-negative orthant is invariant. At \(x_1 = 0\), \(x = x_2 + x_3 \geq \epsilon\), and if \(x_2 > 0\):
\[
\implies \dot{x}_1 = \alpha_1 x \cdot \beta_{22} \frac{x_2^2}{4 x^2} \geq 0,
\]
growth is positive , ensuring \(x_1 \geq 0\). 

\noindent If \(x_2 = 0\), \(\dot{x}_1 = 0\), so \(x_1 = 0\) is invariant. Similarly, for \(x_2 = 0\): $\dot{x}_2 = 0 $
and for \(x_3 = 0\):$$\dot{x}_3 = \alpha_3 x \cdot \frac{1}{4} \beta_{22} \frac{x_2^2}{x^2} \geq 0 $$
For \(y = 0\), \(\implies \dot{y} = 0\). Thus, \(x_1, x_2, x_3, y \geq 0\).

\noindent\paragraph{Boundedness of Total Prey Population}
\noindent  Let \(X = x_1 + x_2 + x_3\). We compute: $\dot{X} = \dot{x}_1 + \dot{x}_2 + \dot{x}_3$. We take up the estimates for the total prey population by analyzing the  term by term boundedness of the vector field $\textbf{f}$ .The growth terms are bounded by:

\[
\alpha_1 x \sum \beta_{ij} \frac{x_i x_j}{x^2} \leq \alpha_1 \beta_{\max} X, \quad 2 \alpha_2 x \sum \beta_{ij} \frac{x_i x_j}{x^2} \leq 2 \alpha_2 \beta_{\max} X, \quad \alpha_3 x \sum \beta_{ij} \frac{x_i x_j}{x^2} \leq \alpha_3 \beta_{\max} X,
\]

\noindent where \(\beta_{\max} = \max_{i,j} |\beta_{ij}|\). Thus, the growth is at most \(C \max_i (\alpha_i) X\), with \(C = 3 \beta_{\max}\). The carrying capacity terms are:

\[
-\frac{1}{K} (\alpha_1 x_1 x + \alpha_2 x_2 x + \alpha_3 x_3 x) = -\frac{X^2}{K} \sum \alpha_i \frac{x_i}{X} \leq -\frac{\min_i (\alpha_i)}{K} X^2.
\]

\noindent The predation, death, and control terms satisfy:
\[
-\sum_{i=1}^3 \frac{x_i m_i x y}{a + x} - \sum_{i=1}^3 d_i x_i - u x_3 \leq -\min_i (d_i) X,
\]
since \(u \geq 0\). Thus adding all the above inequalities on the individual terms :
\[
\dot{X} \leq C \max_i (\alpha_i) X - \frac{\min_i (\alpha_i)}{K} X^2 - \min_i (d_i) X.
\]
From this we get the equilibrium as \(X = \frac{C \max_i (\alpha_i) - \min_i (d_i)}{\min_i (\alpha_i) / K}\), so \(X(t) \leq X_{\max} = \max(X(0), \frac{K [C \max_i (\alpha_i) - \min_i (d_i)]}{\min_i (\alpha_i)})\). If \(X = \epsilon\), growth terms are positive, ensuring \(X \geq \epsilon\).

\paragraph{Boundedness of Predator population }
\noindent Similarly for the predator we have, since \(X \leq X_{\max}\):
\[
\dot{y} = y \left( \frac{m_1 x_1 + m_2 x_2}{a + x} - s \right) - \frac{m_3 x_3 y}{a + x} \leq y \left( \frac{\max(m_1, m_2) X}{a} - s \right).
\]
By assumption, \(\max(m_1, m_2) X_{\max} < s a\), so:
\[
\dot{y} \leq y \left( \frac{\max(m_1, m_2) X_{\max}}{a} - s \right) = -k y,
\]

where \(k = s - \frac{\max(m_1, m_2) X_{\max}}{a} > 0\). Thus, \(y(t) \leq y(0) e^{-k t} + \text{bounded terms}\), ensuring \(y\) is bounded.

\paragraph{Global Existence}
Since \(\mathbf{x}(t)\) is non-negative, \(X \leq X_{\max}\), \(y \leq y_{\max}\), and \(x_1 + x_2 + x_3 \geq \epsilon\), solutions remain in a compact subset of \(D\). By Lipschitz continuity of \(f\), solutions cannot blow up in finite time, so they exist  \(\forall t \geq 0\).

Theorems \ref{thm:existence_uniqueness} and \ref{thm:global} ensure the system is well-posed, enabling further analysis of optimal control strategies.

\subsection{Optimal Control Analysis via Pontryagin’s Maximum Principle}\label{subsec:pmp_analysis}

\noindent Having established the existence and uniqueness of solutions to the controlled population dynamics system (Theorems \ref{thm:existence_uniqueness} and \ref{thm:global}), we formulate and analyze an optimal control problem to suppress the mutant population (\(x_3\)) while maintaining ecological stability. We apply Pontryagin’s Maximum Principle (PMP) to derive necessary conditions for the optimal control, providing a framework for numerical implementation due to the system’s high nonlinearity. We have followed (Ch:4-5 of \cite{kirk} and Ch:2, Sec 5,12,15 in \cite{fleming}) for the methods used in the proofs.

\subsubsection{Problem Formulation}

The optimal control problem is defined as follows:

\begin{itemize}
    \item \textbf{States}: The state vector is \(\mathbf{x}(t) = [x_1(t), x_2(t), x_3(t), y(t)]^T \in D\), where \(D = \{ \mathbf{x} \in \mathbb{R}^4 : x_1, x_2, x_3, y \geq 0, x_1 + x_2 + x_3 \geq \epsilon \}\), representing the wild-type, heterozygous, mutant, and predator populations.
    \item \textbf{Control}: The control input is \(u(t) \in U = [0, u_{\max}]\), where \(u_{\max} > 0\), representing the suppression effort (e.g., culling or treatment) targeting \(x_3\).
    \item \textbf{Dynamics}: The system is governed by \eqref{eq:x1}--\eqref{eq:y}, written as:
    \[
    \dot{\mathbf{x}}(t) = f(\mathbf{x}(t), u(t)), \quad \mathbf{x}(0) = \mathbf{x}_0,
    \]
    where \(f(\mathbf{x}, u) = [f_1(\mathbf{x}), f_2(\mathbf{x}), f_3(\mathbf{x}, u), f_4(\mathbf{x})]^T\), and only \(f_3\) depends on \(u\):
    \[
    f_3(\mathbf{x}, u) = \alpha_3 x \left( \beta_{33} \frac{x_3^2}{x^2} + \beta_{32} \frac{x_3 x_2}{x^2} + \frac{1}{4} \beta_{22} \frac{x_2^2}{x^2} \right) - \frac{\alpha_3}{K} x_3 x - \frac{x_3 m_3 x y}{a + x} - d_3 x_3 - u x_3.
    \]
    \item \textbf{Objective Functional}: The goal is to keep the population of \(x_3\) to near the target small population size, stabilize the ecosystem, and limit control effort. The performance index is:
    \begin{equation}
        J(u) = \int_0^T \left[ q x_3^2 + c u^2 + \epsilon (x_1^2 + x_2^2) \right] dt + p x_3(T)^2, \label{eq:cost}
    \end{equation}
    where \(q, c, p > 0\) are weights penalizing the mutant population and control effort, and \(\epsilon = 10^{-6}\) is a small regularization parameter to ensure numerical stability of \(x_1\) and \(x_2\). The terminal cost \(p x_3(T)^2\) drives \(x_3(T) \to 0\).
\end{itemize}

\subsubsection{Hamiltonian Formulation}

\noindent The Hamiltonian combines the running cost and system dynamics:

\begin{equation}
    \mathcal{H}(\mathbf{x}, u, \bm{\lambda}) = q x_3^2 + c u^2 + \epsilon (x_1^2 + x_2^2) + \sum_{i=1}^4 \lambda_i f_i(\mathbf{x}, u), \label{eq:hamiltonian}
\end{equation}

where \(\bm{\lambda} = [\lambda_1, \lambda_2, \lambda_3, \lambda_4]^T \in \mathbb{R}^4\) is the costate vector. The PMP states that if \(u^*(t)\) is optimal with corresponding state \(\mathbf{x}^*(t)\), there exists \(\bm{\lambda}(t)\) satisfying for almost all \(t \in [0, T]\):

\noindent The state dynamics is given by :
\[
\dot{\mathbf{x}}^*(t) = \frac{\partial \mathcal{H}}{\partial \bm{\lambda}} = f(\mathbf{x}^*(t), u^*(t)), \quad \mathbf{x}^*(0) = \mathbf{x}_0.
\]

The costate dynamics thus becomes :
\[
\dot{\lambda}_i(t) = -\frac{\partial \mathcal{H}}{\partial x_i}(\mathbf{x}^*(t), u^*(t), \bm{\lambda}(t)), \quad i = 1, 2, 3, 4,
\]
with terminal conditions from the terminal cost \(\phi(\mathbf{x}(T)) = p x_3(T)^2\):

\[
\lambda_i(T) = \frac{\partial \phi}{\partial x_i}(\mathbf{x}(T)) = \begin{cases} 
    2 p x_3(T) & \text{if } i = 3, \\
    0 & \text{if } i = 1, 2, 4.
\end{cases}
\]

\noindent Finally we get for the costate equations as :
\begin{align}
        \dot{\lambda}_i= \frac{\partial \mathcal{H}}{\partial x_i} &= \frac{\partial}{\partial x_i} \left[ q x_3^2 + \epsilon (x_1^2 + x_2^2) \right] + \sum_{j=1}^4 \lambda_j \frac{\partial f_j}{\partial x_i}\\
    \mbox{For i=1: \ }    \dot{\lambda}_1 &= -2 \epsilon x_1 - \sum_{j=1}^4 \lambda_j \frac{\partial f_j}{\partial x_1}, \quad \lambda_1(T) = 0\\
    \mbox{For i=2:\ } \dot{\lambda}_2 &= -2 \epsilon x_2 - \sum_{j=1}^4 \lambda_j \frac{\partial f_j}{\partial x_2}, \quad \lambda_2(T) = 0\\
    \mbox{For i=3:\ }:\dot{\lambda}_3 &= -2 q x_3 - \sum_{j=1}^4 \lambda_j \frac{\partial f_j}{\partial x_3}, \quad \lambda_3(T) = 2 p x_3(T)\\
    \mbox{For i=4:\ }\dot{\lambda}_4 &= -\sum_{j=1}^4 \lambda_j \frac{\partial f_j}{\partial y}, \quad \lambda_4(T) = 0       
\end{align}

\noindent \textbf{Minimization Condition}:
\begin{equation}
u^*(t) = \arg \min_{u \in [0, u_{\max}]} \mathcal{H}(\mathbf{x}^*(t), u, \bm{\lambda}(t))
\end{equation}

\noindent To determine the optimal control, we minimize the Hamiltonian \(\mathcal{H}\) with respect to \(u\), keeping all other variables fixed:

\[
\mathcal{H} = q x_3^2 + c u^2 + \epsilon (x_1^2 + x_2^2) + \lambda_1 f_1 + \lambda_2 f_2 + \lambda_3 (f_3 - u x_3) + \lambda_4 f_4.
\]
\noindent The \(u\)-dependent part of the Hamiltonian is:
\[
h(u) = c u^2 - \lambda_3 x_3 u.
\]
\noindent Taking the derivative with respect to \(u\) and setting it to zero gives the unconstrained minimizer:
\[
\frac{\partial h}{\partial u} = 2 c u - \lambda_3 x_3 = 0 \quad \Rightarrow \quad u = \frac{\lambda_3 x_3}{2 c}.
\]
\noindent Projecting this onto the admissible control set \(u \in [0, u_{\max}]\) yields the optimal control:
\begin{equation}
    u^*(t) = \begin{cases} 
        0 & \text{if } \frac{\lambda_3(t) x_3^*(t)}{2 c} < 0, \\
        \frac{\lambda_3(t) x_3^*(t)}{2 c} & \text{if } 0 \leq \frac{\lambda_3(t) x_3^*(t)}{2 c} \leq u_{\max}, \\
        u_{\max} & \text{if } \frac{\lambda_3(t) x_3^*(t)}{2 c} > u_{\max}.
    \end{cases} \label{eq:optimal_control}
\end{equation}

\noindent This control law may exhibit a bang-bang or interior (regular) structure. Bang-bang control arises when the expression \(\frac{\lambda_3(t) x_3^*(t)}{2 c}\) lies outside the admissible bounds, resulting in \(u^*(t) = 0\) or \(u^*(t) = u_{\max}\). When the control lies strictly within bounds and the switching function,
\[
\frac{\partial \mathcal{H}}{\partial u} = 2 c u - \lambda_3 x_3,
\]
vanishes identically over a nontrivial time interval, the system enters a singular arc. In such cases, higher-order necessary conditions—such as the vanishing of time derivatives of the switching function—must be evaluated. Due to the nonlinearity of the system, such conditions are generally resolved numerically.

\subsection{Numerical simulation results}\label{subsec:numerical_control}
\begin{figure*}[htbp]
    \centering  
    \includegraphics[scale=0.3]{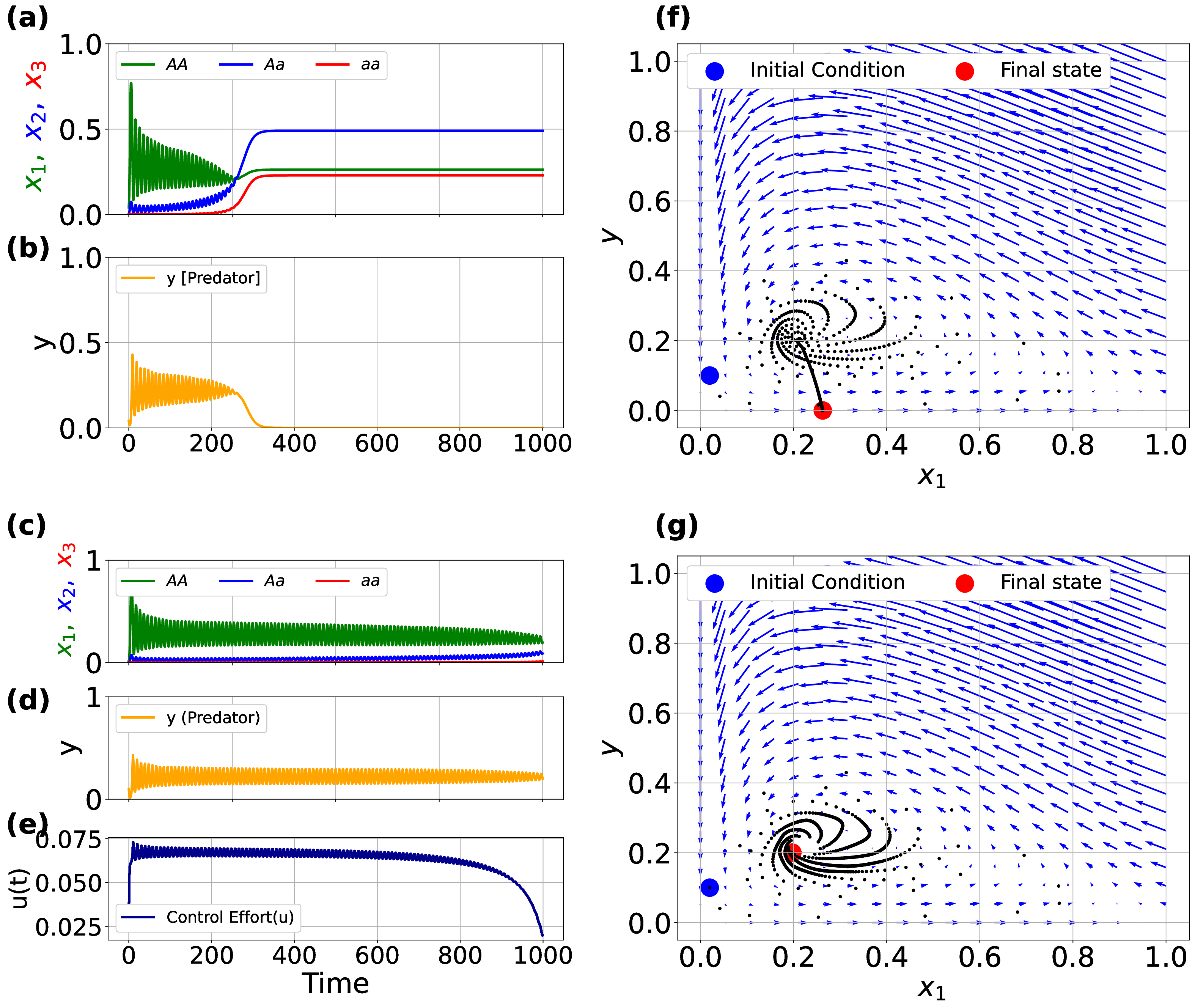}
    \caption{For the parameter values as in main-text and the case $s=1.1,\  d=0.021$ Figures (a) and (b) show the uncontrolled temporal dynamics of prey populations (\(x_1, x_2, x_3\)) and predator population (\(y\)), respectively, with oscillatory amplitudes decaying toward equilibrium over time. Subplots (c) and (d) illustrate the controlled prey and predator populations, stabilized by a diminishing control input (e) (\(u(t)\)), which suppresses oscillations. The phase portraits in (f) and (g) (uncontrolled vs. controlled systems) highlight the transition from cyclic behavior (spiraling trajectories in \(x_1\)-\(y\) plane) to convergence to a stable equilibrium ("Final State"). Time-series plots (a–d) confirm progressive damping of predator-prey cycles, while (e) demonstrates the phased reduction of the control effort. Vector fields (f) and (g) explicitly show the system's evolution from initial oscillatory conditions to stabilized dynamics under control.}
    \label{fig1}
\end{figure*}

\begin{figure*}[ht]
    \centering
    \includegraphics[width=0.95\textwidth]{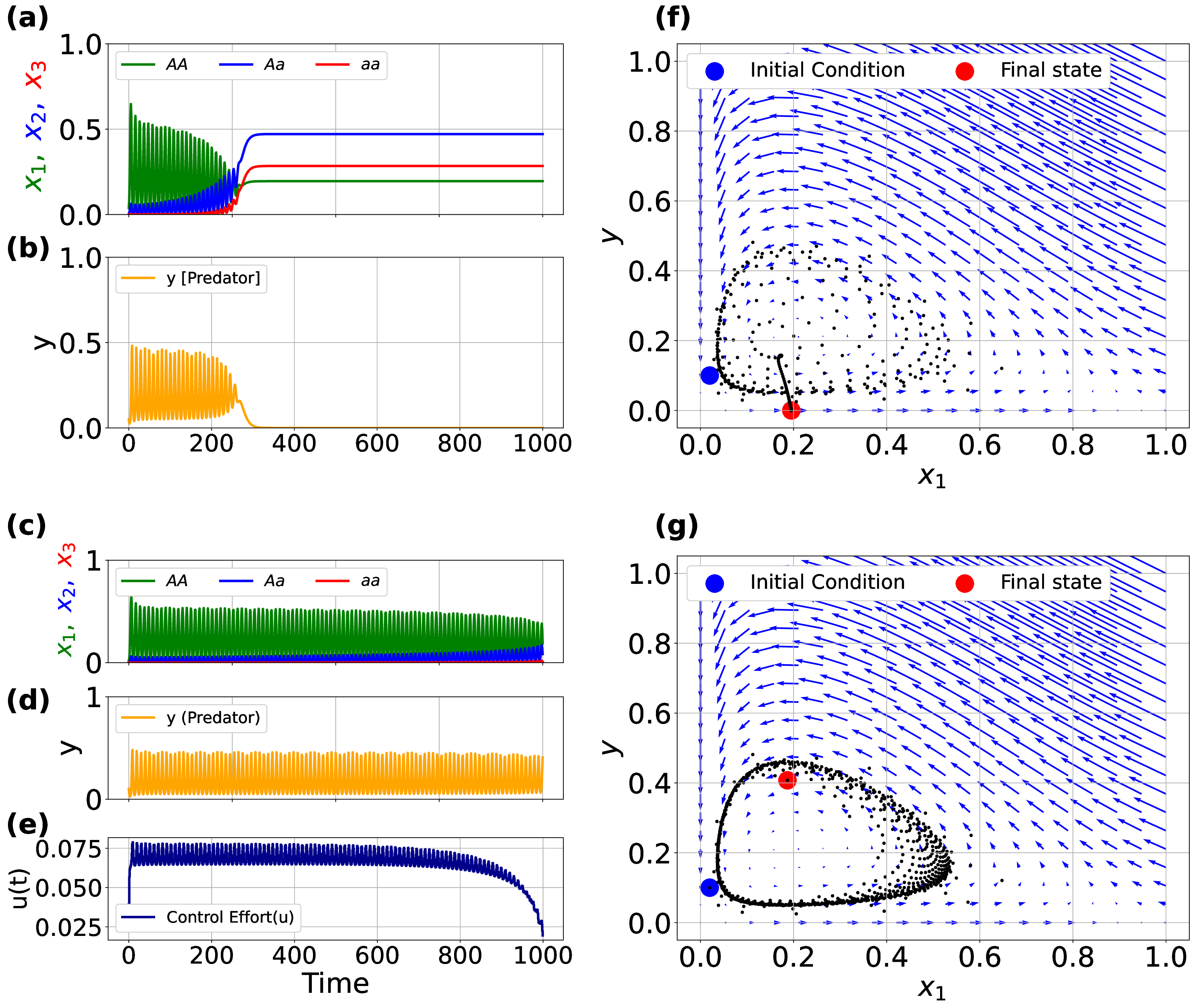}
    \caption{ For the parameter values as in main-text and the case $s=0.891,\  d=0.0591$. Figures (a) and (b) depict the uncontrolled temporal dynamics of prey populations (\(x_1, x_2, x_3\)) and predator population (\(y\)), respectively. Subplots (c), (d), and (e) illustrate the controlled dynamics of prey, predator, and the control input (\(u(t)\) applied to \(x_3\)), respectively. Subplots (f) and (g) present vector fields in the \(x_1\)-\(y\) plane with  $x_2, x_3$ fixed: the uncontrolled system (f) exhibits trajectories spiraling toward extinction (\(y \approx 0\)), whereas the controlled system (g) reveals a stable limit cycle, signifying sustained oscillations. (e) shows the control signal\(u(t)\); persistent oscillations maintain \(y > 0\) as desired.}
    \label{fig2}
\end{figure*}

\noindent The numerical results we present here are for the simulations, where, we have selected the parameter vales to be in the oscillatory range for this system. We, then, numerically solved model system of equations with the optimal control problem formulated in Section \ref{subsec:pmp_analysis} to validate the PMP conditions and demonstrate ecological rescue of the predator population (\(y\)) by suppressing the mutant population (\(x_3\)). Due to the system’s nonlinearity, we solve the resulting two-point boundary value problem (TPBVP), comparing controlled and uncontrolled dynamics to assess the control’s efficacy. The model theoretically examines the impact of genetic variation on species dynamics, using biologically valid parameters : \(\alpha_i =  1.2\), \(\beta_{ij} = 1\), \(K = 1\), \(m_1 = m_2 = 2.5\), \(m_3 = 2.25\), \(a = 0.37\). Initial conditions are \(\mathbf{x}(0) = [0.02, 0.002, 0.00001, 0.1]^T\), reflecting a small initial mutant population. We make two cases one for \(s = 1.1\), and \(d_1 = d_2 = d_3 = 0.021\) and second one as \(s = 0.891\), and \(d_1 = d_2 = d_3 = 0.0591\). We have plotted the dynamics in Fig. \ref{fig1} and in Fig. \ref{fig2}  respectively.
\noindent The time horizon \([0, 1000]\) is discretized into \(N = 1000\) intervals (\(\Delta t = 1\)). The control \(u(t) \in [0, 1]\) allows a full range of suppression efforts. The performance index is:
\[
J(u) = \int_0^T \left[ q (x_3^{\text{target}} - x_3)^2 + r x_3^2 + c u^2 + \epsilon (x_1^2 + x_2^2) \right] dt + p (x_3^{\text{target}} - x_3(T))^2,
\]
with \(q = 1.0\), \(r = 3.0\), \(c = 0.1\), \(\epsilon = 10^{-6}\), \(p = 10\), and \(x_3^{\text{target}} = 0.001\). The terms \(q (x_3^{\text{target}} - x_3)^2 + r x_3^2\) minimize \(x_3\), \(c u^2\) limits control effort, \(\epsilon (x_1^2 + x_2^2)\) ensures numerical stability, and the terminal cost reinforces suppression of \(x_3(T)\).
The simulation uses CasADi’s Opti stack, with the control sequence \(U = [u(t_0), u(t_1), \ldots, u(t_{N-1})]^T \in \mathbb{R}^N\) as decision variables. State trajectories are computed via:
\[
\mathbf{x}(t_{k+1}) = F(\mathbf{x}(t_k), u(t_k)),
\]
where \(F\) is the CVODES integrator solving \eqref{eq:x1}--\eqref{eq:y} over \([t_k, t_{k+1}]\). The cost is approximated as:
\[
J \approx \sum_{k=0}^{N-1} L(\mathbf{x}(t_k), u(t_k)) \Delta t + p (x_3^{\text{target}} - x_3(t_N))^2,
\]
where \(L = q (x_3^{\text{target}} - x_3)^2 + r x_3^2 + c u^2 + \epsilon (x_1^2 + x_2^2)\). Control constraints \(0 \leq u(t_k) \leq 1\) are enforced, and the NLP is solved using IPOPT (tolerance \(10^{-2}\), maximum iterations 500, MUMPS linear solver). The initial control guess is \(u = 0.1\). \\

\noindent The numerical simulations comprehensively illustrate the predator-prey system’s dynamics with and without optimal control, highlighting the ecological rescue of the predator \(y\) under two parameter sets:  Fig. \ref{fig1} for the set \(s = 1.1\), \(d = 0.021\) and Fig.\ref{fig2} for set \(s = 0.891\), \(d = 0.0591\). In the uncontrolled scenario (Fig. \ref{fig1}a, Fig. \ref{fig1}b,  Fig. \ref{fig1}f), the mutant prey population \(x_3\) rapidly proliferates due to its relatively lower capture rate, and due to its harmful nature it significantly increases predation pressure on \(y\). This unchecked growth drives \(y\) toward extinction, as evidenced by the quiver phase plot (Fig. \ref{fig1}f), where trajectories spiral inward to \(y \to 0\), signaling a collapse of predator-prey interactions. To counteract this decline, an optimal control strategy \(u(t)\) is applied, effectively suppressing \(x_3\) to a predefined target threshold while minimizing a cost function. Under control the dynamics are plotted in (Fig. \ref{fig1}c-e,  Fig. \ref{fig1}g), \(x_3\) is regulated near the target, alleviating predation pressure and enabling \(y\) to stabilize above zero in both cases. In Figure \ref{fig1}, the slightly high predator mortality (\(s = 1.1\)) causes \(y\) to lose biomass over time, leading to oscillations in \(x_1\), \(x_2\), and \(y\) that decay in amplitude (Fig. \ref{fig1}g), with the phase plot showing trajectories converging to a small limit cycle, indicative of weakened ecological interactions. The low prey mortality (\(d = 0.021\)) exacerbates this by allowing \(x_1\) and \(x_2\) to grow steadily, further dampening cyclic dynamics. \\

\noindent In contrast, Fig. \ref{fig2}  (\(s = 0.891, d = 0.0591\)) demonstrates a more robust recovery: the reduced \(s\) minimizes \(y\)’s biomass loss, while the higher \(d\) curbs prey overgrowth, fostering larger, sustained oscillations in \(x_1\), \(x_2\), and \(y\) (Fig. \ref{fig2}g), with the phase plot reflecting a healthy limit cycle. Although the control successfully rescues \(y\) in both scenarios, the predator-prey interaction in Figure \ref{fig1}  is notably weaker than in Figure \ref{fig2} , where the favorable mortality rates enhance ecological stability and cyclic vigor. These simulations, computed over \(t \in [0, 1000]\) with \(N = 1000\) intervals, underscore the control’s efficacy in preventing extinction across varying conditions, while highlighting the pivotal role of \(s\) and \(d\) in determining the strength and sustainability of predator-prey interactions.

\section{Discussion}
\label{sec:discussion}

\noindent This study demonstrates the ecological rescue of a predator population (\(y\)) threatened by a mutant population (\(x_3\)) in a four-species nonlinear system. Numerical simulations reveal that uncontrolled dynamics lead to \(x_3\) dominance, predator extinction, and oscillation collapse, while the optimal control, derived via Pontryagin’s Maximum Principle (PMP), suppresses \(x_3\), sustains \(y\), and restores oscillatory dynamics. These findings underscore the importance of targeted interventions to maintain predator-prey cycles, critical for ecological stability. The proofs of local existence/uniqueness (Theorem \ref{thm:existence_uniqueness}) and global boundedness (Theorem \ref{thm:global}) ensure the model’s mathematical robustness, extending traditional ecological models with genetic and control elements. However, constant parameters and fixed control bounds (\([0.05, 0.1]\)) simplify real-world variability and resource constraints. Numerical reliance due to nonlinearity poses computational challenges. Future work could incorporate stochasticity, spatial dynamics, or multi-input controls, and validate the model with field data from invasive species scenarios. This framework informs conservation strategies, offering a mathematically rigorous approach to managing ecological disruptions and preserving biodiversity.

\section{Conclusion}
\label{sec:conclusion}

\noindent This work presents a robust framework for ecological rescue by controlling a mutant population in a four-species nonlinear system. Through a nonlinear ODE model, we prove local and global existence of solutions, ensuring biological realism. Applying PMP, we derive an optimal control that minimizes the mutant population, rescuing the predator and restoring ecological oscillations, as validated by numerical simulations. The contrast between uncontrolled (predator extinction, oscillation collapse) and controlled (rescue, oscillation restoration) dynamics highlights the control’s efficacy. This study bridges applied mathematics and ecology, providing a versatile tool for managing invasive or mutant populations. Its applications extend to conservation, invasive species control, and ecosystem restoration, with potential in microbial or disease dynamics. Future research could explore spatial models or empirical validation to enhance practical impact. This framework advances our ability to design precise interventions for sustainable ecological management.\\

\appendix

\noindent\textbf{Author Contributions:} Conceptualization, P.M., S.K., R.K.B.S.; Analysis and numerical simulations, P.M. , S.K. \& S.C.C.V.; Writing—original draft preparation P.M., S.K., R.K.B.S.; Writing—review and editing, P.M., S.K., S.C.C.V \& R.K.B.S.; Supervision, R.K.B.S.; All authors have read and approved the manuscript.\\

\noindent\textbf{Acknowledgment:} P.M. acknowledges CSIR for Senior Research Fellowship funding. S.K. acknowledges ICMR for Senior Research Fellowship funding. S.C.C.V. acknowledges CSIR, UGC for Junior research Fellowship funding . R.K.B.S.hg acknowledges JNU and DBT BIC for facilities.\\

\noindent\textbf{Data Availability Statement:} No data was used in this article.\\

\noindent\textbf{Conflicts of Interest:} The authors declare no conflicts of interest.



\begin{thebibliography}{}



\bibitem{cadotte}Cadotte, M. W., Mcmahon, S. M., \& Fukami, T. (Eds.). (2006). Conceptual ecology and invasion biology: reciprocal approaches to nature (Vol. 1). Springer Science \& Business Media.\url{https://doi.org/10.1007/1-4020-4925-0}

\bibitem{courchamp} Courchamp, F., Caut, S. (2006). Use of biological invasions and their control to study the dynamics of interacting populations. In: Cadotte, M.W., Mcmahon, S.M., Fukami, T. (eds) Conceptual Ecology and Invasion Biology: Reciprocal Approaches to Nature., vol 1. Springer, Dordrecht. \url{https://doi.org/10.1007/1-4020-4925-0_11}


\bibitem{braselton1}Waltman, P., Braselton, J., Braselton, L. (2002). A mathematical model of a biological arms race with a dangerous prey. Journal of theoretical biology, 218(1), 55-70. \url{https://doi.org/10.1006/jtbi.2002.3057}


\bibitem{sun}Sun, K., Zhang, T., \& Tian, Y. (2017). Dynamics analysis and control optimization of a pest management predator–prey model with an integrated control strategy. Applied Mathematics and Computation, 292, 253-271.\url{https://doi.org/10.1016/j.amc.2016.07.046}


\bibitem{kirk}Kirk, D. E. (2004). Optimal control theory: an introduction. Courier Corporation.

\bibitem{lenhart}Lenhart, S., \& Workman, J. T. (2007). Optimal control applied to biological models. Chapman and Hall/CRC.\url{https://doi.org/10.1201/9781420011418}


\bibitem{casadi}Andersson, J.A.E., Gillis, J., Horn, G. et al. (2019) CasADi: a software framework for nonlinear optimization and optimal control. Math. Prog. Comp. 11, 1–36 . \url{https://doi.org/10.1007/s12532-018-0139-4}



\bibitem{brodie1}Brodie III, E. D., \& Brodie Jr, E. D. (1990). Tetrodotoxin resistance in garter snakes: an evolutionary response of predators to dangerous prey. Evolution, 44(3), 651-659.\url{https://doi.org/10.1111/j.1558-5646.1990.tb05945.x}

\bibitem{brodie2} Williams, B. L., Brodie Jr, E. D., \& Brodie III, E. D. (2003). Coevolution of deadly toxins and predator resistance: self-assessment of resistance by garter snakes leads to behavioral rejection of toxic newt prey. Herpetologica, 59(2), 155-163.\url{https://www.jstor.org/stable/3893352}

\bibitem{freedman}Freedman, H. I., \& Waltman, P. (1978). Predator influence on the growth of a population with three genotypes. Journal of Mathematical Biology, 6(4), 367-374. \url{ https://doi.org/10.1007/BF0246300}
\bibitem{drossel}Thiel, T., Brechtel, A., Brückner, A., Heethoff, M., \& Drossel, B. (2019). The effect of reservoir-based chemical defense on predator-prey dynamics. Theoretical Ecology, 12, 365-378.\url{https://link.springer.com/article/10.1007/s12080-018-0402-3}

\bibitem{abrams}Abrams, P. A. (2000). The evolution of predator-prey interactions: theory and evidence. Annual Review of Ecology and Systematics, 31(1), 79-105.\url{https://doi.org/10.1146/annurev.ecolsys.31.1.79}


\bibitem{schoener}Schoener, T. W. (2011). The newest synthesis: understanding the interplay of evolutionary and ecological dynamics. science, 331(6016), 426-429.\url{https://doi.org/10.1126/science.1193954}


\bibitem{montroll}N. S. Goel, S. C. Maitra and E. W. Montroll.(1971)  On the Volterra and Other Nonlinear Models of Interacting Populations. Rev. Mod. Phys. 43, 231–276 \url{https://doi.org/10.1103/RevModPhys.43.231}.


\bibitem{May1975}May, R. M.,  Leonard, W. J. (1975). Nonlinear Aspects of Competition Between Three Species. SIAM Journal on Applied Mathematics, 29(2), 243–253. \url{https://doi.org/10.1137/0129022}

\bibitem{Maier2013}Maier R. S. (2013) The integration of three-dimensional Lotka–Volterra systems. Proc. R. Soc. A 469 20120693
\url{http://doi.org/10.1098/rspa.2012.0693}

\bibitem{koth}Koth, M.,  Kemler, F. (1986). A one locus-two allele selection model admitting stable limit cycles. Journal of theoretical biology, 122(3), 263-267.\url{https://doi.org/10.1016/S0022-5193(86)80119-9}

\bibitem{akin}Akin, E. (1982). Cycling in simple genetic systems. Journal of Mathematical Biology, 13(3), 305-324.\url{https://link.springer.com/article/10.1007/BF00276066}

\bibitem{cleve}Van Cleve, J.,  Feldman, M. W. (2008). Stable long-period cycling and complex dynamics in a single-locus fertility model with genomic imprinting. Journal of mathematical biology, 57, 243-264.\url{https://doi.org/10.1007/s00285-008-0156-4}

\bibitem{braselton2}Braselton, J., Abell, M., Braselton, L. (2005). Selective mating in a continuous model of epistasis. Applied mathematics and computation, 171(1), 225-241. \url{https://doi.org/10.1016/j.amc.2005.01.059}



\bibitem{conte} Conte, G., Moog, C. H.,  Perdon, A. M. (1999). Nonlinear control systems: An algebraic setting\url{https://doi.org/10.1007/BFb0109765}.


\bibitem{brockett1} Brockett, R. W. (2014). The early days of geometric nonlinear control. Automatica, 50(9), 2203-2224\url{https://doi.org/10.1016/j.automatica.2014.06.010}.

\bibitem{brockett2} Brockett, R. W. (2015). Finite dimensional linear systems. Society for Industrial and Applied Mathematics\url{https://epubs.siam.org/doi/pdf/10.1137/1.9781611973884.bm}.

\bibitem{brockett3} Brockett, R. W. (1973). Lie algebras and Lie groups in control theory. In Geometric methods in system theory (pp. 43-82). Springer, Dordrecht\url{https://doi.org/10.1007/978-94-010-2675-8_2}.

\bibitem{neijmeier} Nijmeijer, H., Van der Schaft, A. (1990). Nonlinear dynamical control systems (Vol. 175). New York: Springer-verlag\url{ https://doi.org/10.1007/978-1-4757-2101-0}.


\bibitem{fleming}Fleming, W. H., \& Rishel, R. W. (2012). Deterministic and stochastic optimal control (Vol. 1). Springer Science \& Business Media.\url{https://link.springer.com/book/10.1007/978-1-4612-6380-7}


\bibitem{strogatz}Strogatz, S. H. (2024). Nonlinear dynamics and chaos: with applications to physics, biology, chemistry, and engineering. Chapman and Hall/CRC.


\bibitem{perko}Perko, L. (2013). Differential equations and dynamical systems (Vol. 7). Springer Science \& Business Media.

\end{thebibliography}
\end{document}